\documentclass[usenatbib]{mn2e}
\newcommand{\bib}{\bibitem[\protect\citeauthoryear}
\usepackage{epsf}
\usepackage[T1]{fontenc}
\usepackage{ae,aecompl}
\begin{document}
\title[Observations of 3C\,31]{Multifrequency VLA observations of the FR\,I
radio galaxy 3C\,31: morphology, spectrum and magnetic field}
\author[R.A. Laing et al.]
   {R.A. Laing \thanks{E-mail: rlaing@eso.org}$^{1}$, A.H. Bridle $^2$,
    P. Parma $^3$, 
    L. Feretti $^{3}$, G. Giovannini $^{3,4}$, \newauthor M. Murgia$^{3,5}$,
    R.A. Perley$^6$\\
    $^1$ European Southern Observatory, Karl-Schwarzschild-Stra\ss e 2, D-85748 
    Garching-bei-M\"unchen, Germany \\ 
    $^2$ National Radio Astronomy Observatory, Edgemont Road, Charlottesville,
    VA 22903-2475, U.S.A. \\
    $^3$ INAF - Istituto di Radioastronomia, via Gobetti 101, I-40129 Bologna,
    Italy \\
    $^4$ Dipartimento di Astronomia, Universit\`{a} di Bologna, 
     Via Ranzani,1 , I-40127 Bologna, Italy \\
    $^5$ INAF - Osservatorio Astronomico di Cagliari, Loc. Poggio dei Pini, Strada 54,
    I-09012 Capoterra (CA), Italy \\   
    $^6$ National Radio Astronomy Observatory, P.O. Box O, Socorro, NM
    87801, U.S.A.
    }

\date{Received }
\maketitle

\begin{abstract}
We present high-quality Very Large Array (VLA) images of the Fanaroff \& Riley
Class I radio galaxy 3C\,31 in the frequency range 1365 to 8440\,MHz with
angular resolutions from 0.25 to 40 arcsec.  Our new images reveal complex, well
resolved filamentary substructure in the radio jets and tails. We also use these
images to explore the spectral structure of 3C\,31 on large and small scales.

We infer the apparent magnetic field structure by correcting for Faraday
rotation. Some of the intensity substructure in the jets is clearly related to
structure in their apparent magnetic field: there are arcs of emission where the
degree of linear polarization increases, with the apparent magnetic field
parallel to the ridges of the arcs.

The spectra of the jets between 1365 and 8440\,MHz are consistent with power
laws within 60\,arcsec of the nucleus.  The spectral indices, $\alpha$ (flux
density $\propto \nu^{-\alpha}$) are significantly steeper ($\alpha = 0.62$)
within $\approx$7\,arcsec of the nucleus than between 7 and 50\,arcsec ($\alpha
= $ 0.52 -- 0.57).  The spectra of the arcs and of the jet edges are also
slightly flatter than the average for their surroundings.  At larger distances,
the jets are clearly delimited from surrounding larger-scale emission both by
their flatter radio spectra and by sharp brightness gradients.

The spectral index of 0.62 in the first 7\,arcsec of 3C\,31's jets is very close
to that found in other FR\,I galaxies where their jets first brighten in the
radio and where X-ray synchrotron emission is most prominent. Farther from the
nucleus, where the spectra flatten, X-ray emission is fainter relative to the
radio. The brightest X-ray emission from FR\,I jets is therefore not associated
with the flattest radio spectra, but with a particle-acceleration process whose
characteristic energy index is $2\alpha+1 = 2.24$. The spectral flattening with
distance from the nucleus occurs where our relativistic jet models require
deceleration, and the flatter-spectra at the jet edges may be associated with
transverse velocity shear.

\end{abstract}


\section{Introduction}
\label{Introduction}

The radio galaxy 3C\,31 is a nearby member of the low-luminosity FR\,I
\citep{FR74} class. Its prominent, symmetrical twin radio jets were discovered
by \citet{Burch77} and imaged at higher resolution by \citet{Fom80}. The jets
are asymmetric on scales up to $\approx$10\,kpc and only the brighter jet is
detected on parsec scales \citep{Lara97}.  More diffuse tails of emission
\citep{Strom83,Jaegers} extend for $\approx$20\,arcmin on either side of the
nucleus.  The source shows strong linear polarization and a pronounced
depolarization asymmetry \citep{Burch79,Strom83}.

The detection of the brighter jet at optical wavelengths by \citet*{BvBM}
was not confirmed by \citet{FB} or \citet{Keel}, but \citet{Croston}
presented  evidence for associated optical emission between 5
and 10\,arcsec from the nucleus.  X-ray emission from this jet was found by
\citet{Hard02}.

3C\,31 is identified with the bright elliptical galaxy NGC\,383, which we
take to have a redshift of 0.0169 (the mean value from
\citealt{Smith2000}, \citealt*{HVG} and \citealt{RC3}).  NGC\,383 is the
brightest member of a rich group of galaxies \citep{Arp66,Zwi68} and, in
common with other nearby radio galaxies, has a dusty nucleus
\citep{Mar99}.  Hot gas on group and galactic scales has been detected by
X-ray imaging \citep{KB99,Hard02}.

Deep 8440-MHz Very Large Array (VLA) images of 3C\,31 at resolutions of 0.75 and
0.25\,arcsec were made by \citet{LB02a}. They showed that the total intensity
and linear polarization within 30\,arcsec of the nucleus can be modelled on the
assumption that the jets are intrinsically symmetrical, axisymmetric,
relativistic and decelerating. They also assumed that the jet flows are
stationary, in the sense that they are characterized by the same global
parameters at all times. By fitting parameterized models of the velocity field,
emissivity and field-ordering, they were able to deduce that the angle to the
line of sight is $\approx$52$^\circ$, to quantify the deceleration and to
establish that transverse velocity gradients must be present in the jets.
\citet{LB02b} used this kinematic model and a description of the external gas
pressure and density from \citet{Hard02} to make a conservation-law
analysis. They showed that the inferred velocity field is consistent with jet
deceleration by entrainment of thermal matter and deduced the energy flux and
the variations of pressure, density and entrainment rate with distance from the
nucleus. \citet{LB04} examined adiabatic models for the jets in 3C\,31: these
models predict too steep a brightness decline along the jets for plausible
variations of the jet velocity and fail to reproduce the observed magnetic-field
structures.

In this paper we describe VLA observations of 3C\,31 
obtained under NRAO observing proposals AF236 and AL405
at frequencies between 1365 and 8440\,MHz and resolutions of 0.25 -- 
40\,arcsec (the 8440-MHz observations are those presented by 
\citealt{LB02a}). We then discuss three aspects of the 
astrophysics of 3C\,31 which complement the treatment of its inner 
jet dynamics by \citet{LB02a,LB02b,LB04} as follows.
\begin{enumerate}
\item {\em Arcs and filaments.} The jets in 3C\,31 contain discrete features --
arcs and non-axisymmetric knots -- visible in both total and linearly polarized
intensity. Such features, and filamentation in general, are common in
extragalactic synchrotron sources when observed with sufficiently high
resolution and good spatial frequency coverage.  In particular, \citet{LCBH06}
found similar arc-like structure in the twin-jet source 3C\,296 and suggested
that systematic differences between the arcs in the main (brighter) and
counter-jets could also be a manifestation of differential relativistic
aberration.  In 3C\,31, we study how the intensity gradients in such features
are related to their magnetic-field structures.
\item {\em Large-scale spectral structure.} 3C\,31 belongs to the sub-class of
FR\,I sources with tails of diffuse emission extending away from the nucleus (as
opposed to lobes with well-defined outer edges).  As in other members of the
class \citep{Parma99}, its radio spectrum steepens with distance from the
nucleus \citep{Burch77,Strom83,And92}. Analyses of such spectral variations from
low-resolution observations (as made for 3C\,31 by \citealt{And92}) interpreted
them as the results of radiative losses on an energy spectrum initially of
power-law form and derived corresponding time-scales, which were identified with
dynamical ages. More recently, high-resolution observations \citep{KSR97,KS99}
have shown that spectral structures of a number of FR\,I sources are better
described as a flat-spectrum `spine' surrounded by a steeper-spectrum `sheath'
rather than as a single component whose spectrum varies smoothly with distance
from the nucleus.  In 3C\,31, we have good spatial resolution over a range of
frequencies and can examine the spectral variations in more detail.
\item {\em Jet base spectra, X-ray emission and particle acceleration
  processes.} We have recently shown that accurate imaging of radio
  spectra in FR\,I jet bases can reveal correlations between spectral index,
  X-ray emission and the jet kinematics as inferred from our relativistic models 
  \citep{LCCB06,LCBH06}. The shortness of the synchrotron lifetimes for
  electrons radiating in the X-ray band requires them to be accelerated in situ,
  so there must be a direct connection between the radio spectral index and the
  particle acceleration process. For 3C\,31 we can determine the radio spectrum
  very accurately at high resolution and examine the connection with X-ray
  emission and kinematics in greater detail.  
 \end{enumerate}

The observations and their reduction are described in Section~\ref{Obs-Red} and
the images are presented in Section~\ref{Images}. Discrete features in total
intensity and the apparent magnetic-field structure are discussed in
Section~\ref{Morph} and variations in radio spectrum across the source are
described in Section~\ref{Spectrum}. Section~\ref{Conclusions} summarizes our
main conclusions. Faraday rotation and depolarization of the polarized emission
are analysed in a companion paper \citep{lb08b} which discusses the structure of
the magnetic field in the foreground screen that is responsible for the Faraday
rotation measure (RM) fluctuations.

We assume a concordance cosmology with a Hubble Constant $H_0$ = 70\,km
s$^{-1}$\,Mpc$^{-1}$, $\Omega_\Lambda = 0.7$ and $\Omega_M = 0.3$. At the
redshift of NGC\,383 (z = 0.0169), this gives a linear scale of
0.344\,kpc/arcsec. We also adopt the sign convention $S(\nu) \propto
\nu^{-\alpha}$ for the spectral index, $\alpha$.

\section{Observations and Data Reduction}
\label{Obs-Red}

\subsection{Observations}
\label{Obs}

The observations were made with the NRAO VLA at frequencies between 1365 and
8440\,MHz and in all four configurations of the array. A journal of observations
is given in Table~\ref{Obstable}.  The flux-density scale was set using
observations of 3C\,48 or 3C\,286 and the zero-point of {\bf E}-vector position
angle with reference to 3C\,138 or 3C\,286.  A single pointing centre
(coincident with the galaxy nucleus) was used for all observations.

\begin{table}
\caption{Journal of observations. The columns are: (1) VLA
configuration, (2) centre frequency, (3) bandwidth, (4) date of
observation and (5) integration time. \label{Obstable}}
\begin{minipage}{120mm}
\begin{tabular}{llrlr}
\hline
&&&&\\
Conf & $\nu$ & $\Delta\nu$ & Date & t \\
     & MHz   &   MHz       &      & min \\
&&&&\\
\hline
&&&&\\
A   & 8460\footnote{The slight difference in centre frequency used for the\\
A-configuration observations has no measurable effect\\ on the results and
we refer throughout to the\\ `8440-MHz observations'.}&100.00 & 1996 Nov 12   &606 \\
    & 1636 &6.25   & 1995 Apr 28   &192 \\
    & 1485 &6.25   & 1995 Apr 28   &192 \\
    & 1435 &6.25   & 1995 Apr 28   &192 \\
    & 1365 &6.25   & 1995 Apr 28   &192 \\
A/B\footnote{Most antennas were in the B configuration.} & 8440 &100.00 & 1994 Jun 6    &406 \\
B   & 8440 &100.00 & 1994 Jun 14   &412 \\
    & 4985 &50.00  & 1993 Feb 25   &535 \\
    & 1636 &25.00  & 1994 Jun 13   &192 \\
    & 1485 &25.00  & 1994 Jun 13   &192 \\
    & 1435 &25.00  & 1994 Jun 13   &192 \\
    & 1365 &25.00  & 1994 Jun 13   &192 \\
C   & 8440 &100.00 & 1994 Dec 4    &242 \\
    & 4985 &50.00  & 1994 Dec 4    &165 \\
    & 1636 &25.00  & 1994 Nov 26   &140 \\
    & 1485 &25.00  & 1994 Nov 26   &140 \\
    & 1435 &25.00  & 1994 Nov 26   &139 \\
    & 1365 &25.00  & 1994 Nov 26   &139 \\
D   & 8440 &100.00 & 1995 Apr 28   &69  \\
    & 4985 &50.00  & 1995 Apr 28   &69  \\
    & 1636 &25.00  & 1995 Apr 28   &61  \\
    & 1485 &25.00  & 1995 Apr 28   &61  \\
    & 1435 &25.00  & 1995 Apr 28   &69  \\
    & 1365 &25.00  & 1995 Apr 28   &69  \\
&&&&\\
\hline
\end{tabular}
\end{minipage}
\end{table}

\begin{table}
\caption{Parameters of the images.\label{Image-table}}
\begin{tabular}{rrrr}
\hline
&&&\\
$\nu$ & FWHM &\multicolumn{2}{c}{$\sigma$ / $\mu$Jy beam$^{-1}$} \\
MHz   & arcsec & I  & QU \\
&&&\\
8440  & 0.25  &   6 & 6   \\
8440  & 0.75  &   7 & 6   \\
&&&\\
8440  & 1.50  &  11 & 6   \\
4985  & 1.50  &  12 & 12  \\
1636  & 1.50  &  72 & 64  \\
1485  & 1.50  &  73 & 66  \\
1435  & 1.50  &  78 & 68  \\
1365  & 1.50  &  76 & 62  \\
Mean L-band & 1.50 & 34 &$-$\\
&&&\\
8440  & 5.50  &  37 & 12  \\
4985  & 5.50  &  26 & 16  \\
1636  & 5.50  &  61 & 22  \\
1485  & 5.50  &  52 & 23  \\
1435  & 5.50  &  68 & 25  \\
1365  & 5.50  &  68 & 23  \\
Mean L-band & 5.50 & 39 &$-$ \\
&&&\\
Mean L-band & 40.00 & 100  &$-$\\
&&&\\
\hline
\end{tabular}
\end{table}

\begin{figure*}
\epsfxsize=17cm
\epsffile{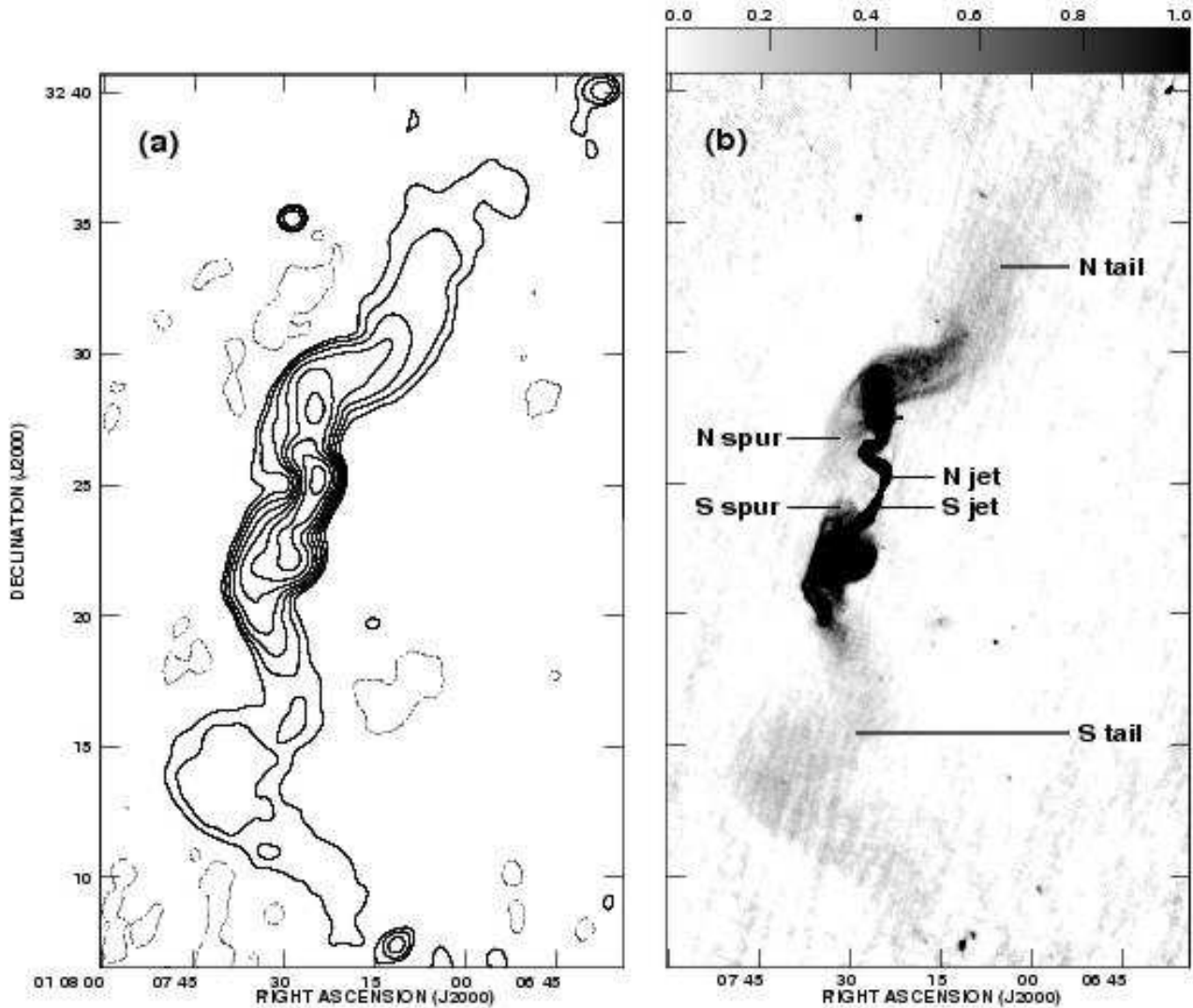}
\caption{Low-resolution, mean L-band images (Section~\ref{Red}) optimized to
show the extent of the observed large-scale structure. (a) Contours at
40\,arcsec FWHM. The levels are $-1$, 1, 2, 4, 8, 16, 32, 64, 128 $\times$
3.5\,mJy beam$^{-1}$. (b) Grey-scale at 5.5\,arcsec FWHM.  The displayed range
is 0 -- 1\,mJy beam$^{-1}$. Key regions are labelled, as discussed in the text.
Note that the primary beam correction leads to increased noise at the edges of
the field and that the low-level stripes in the tails (panel b) are
artefacts. Chromatic aberration is large at the edges of panel (b), so images of
point-like  sources are smeared radially by a factor $\approx$3 and
reduced in peak flux density by the same amount. The effect on the brightness
distribution of 3C\,31, which is fully resolved on these scales, is slight
(Section~\ref{Red}). \label{Ideep}}
\end{figure*}

\subsection{Data reduction}
\label{Red}

Calibration and imaging of the VLA data were done using the {\sc aips} software
package and followed standard procedures with one exception: we used the
iterative technique described by \citet{LB02a} in order to correct for core
variations and amplitude errors when combining data from different
configurations at 4985 and 8440\,MHz. For the lower-frequency observations, core
variability was below the level we could measure and no corrections were made.

We used both {\sc clean} and maximum entropy deconvolution methods for the
total-intensity images and compared the results (in the latter case, we
subtracted the core before deconvolution and restored it again afterwards). In
all cases, the differences between the images (after convolution with the same
Gaussian restoring beam) were less than the larger of 3$\sigma_I$
(Table~\ref{Image-table}) and 0.01$I$. The mean levels were consistent to $\ll
\sigma_I$ and the total flux densities differed by $<$1\%.  As usual, the maximum
entropy images of extended, uniform brightness distributions were locally
smoother and showed little of the patchy structure or high-frequency ripples
characteristic of the {\sc clean} algorithm.  We therefore show the
maximum-entropy $I$ images at resolutions of $\leq$1.5\,arcsec. All quantitative
measurements have also been checked on the {\sc clean} images.

\begin{figure*}
\epsfxsize=17cm
\epsffile{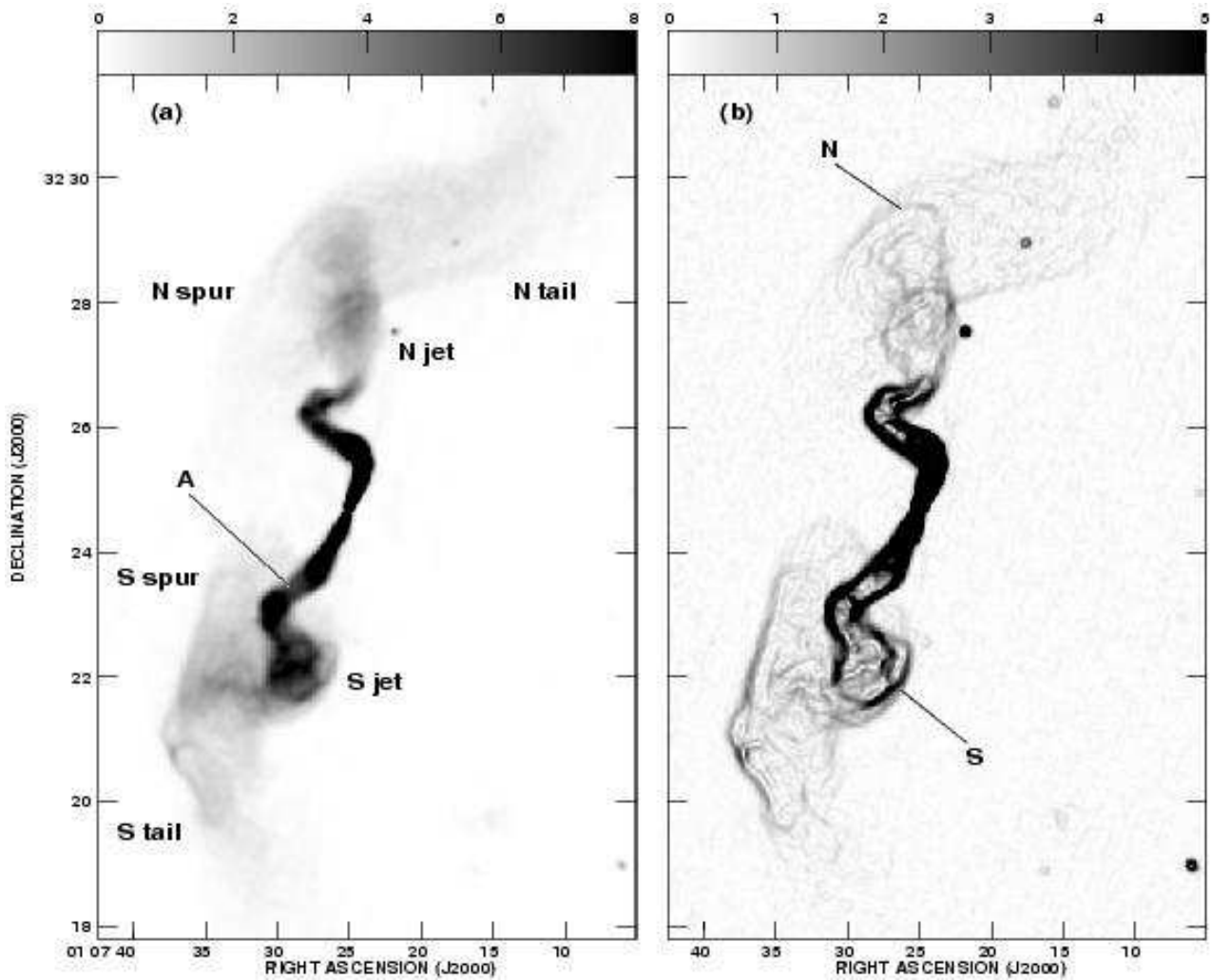}
\caption{Grey-scale representations of the mean L-band image of 3C31
  (Section~\ref{Red}) at a resolution of 5.5 arcsec FWHM.  (a) Total
  intensity. The grey-scale range runs from 0 -- 8 mJy beam$^{-1}$.  (b) A
  Sobel-filtered version of the same image, normalized by the total intensity
  (see text).\label{ISobel5.5}}
\end{figure*}

\begin{figure*}
\epsfxsize=17cm 
\epsffile{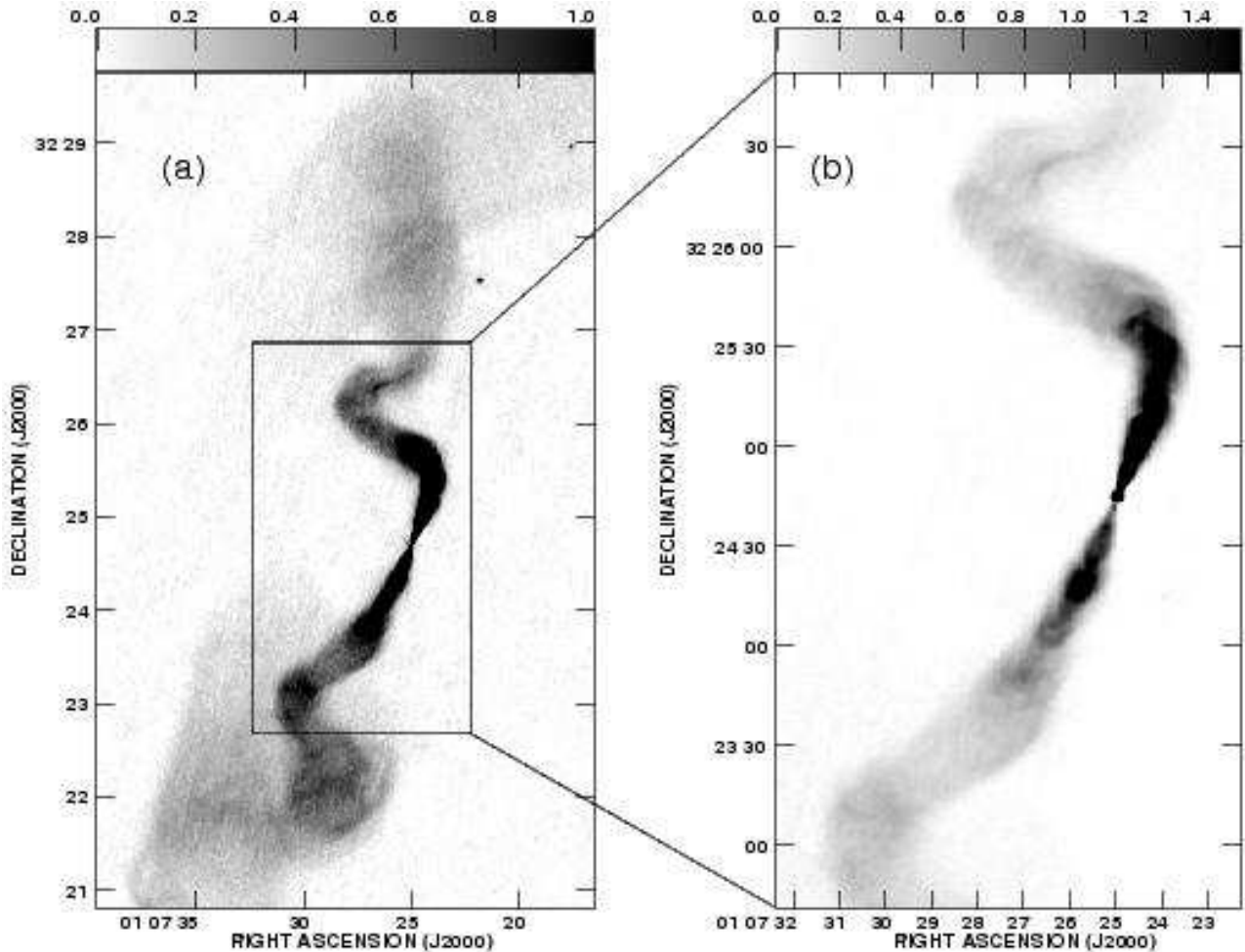}
\caption{Grey-scale images of 3C31 at a resolution of 1.5 arcsec FWHM.  (a) Mean
L-band maximum-entropy deconvolution (Section~\ref{Red}), 0 -- 1\,mJy
beam$^{-1}$. (b) 4985\,MHz maximum entropy deconvolution, 0 -- 1.5\,mJy
beam$^{-1}$.  The area covered by (b) is shown by the box on (a).
\label{I1.5}}
\end{figure*}

\begin{figure*}
\epsfxsize=17cm
\epsffile{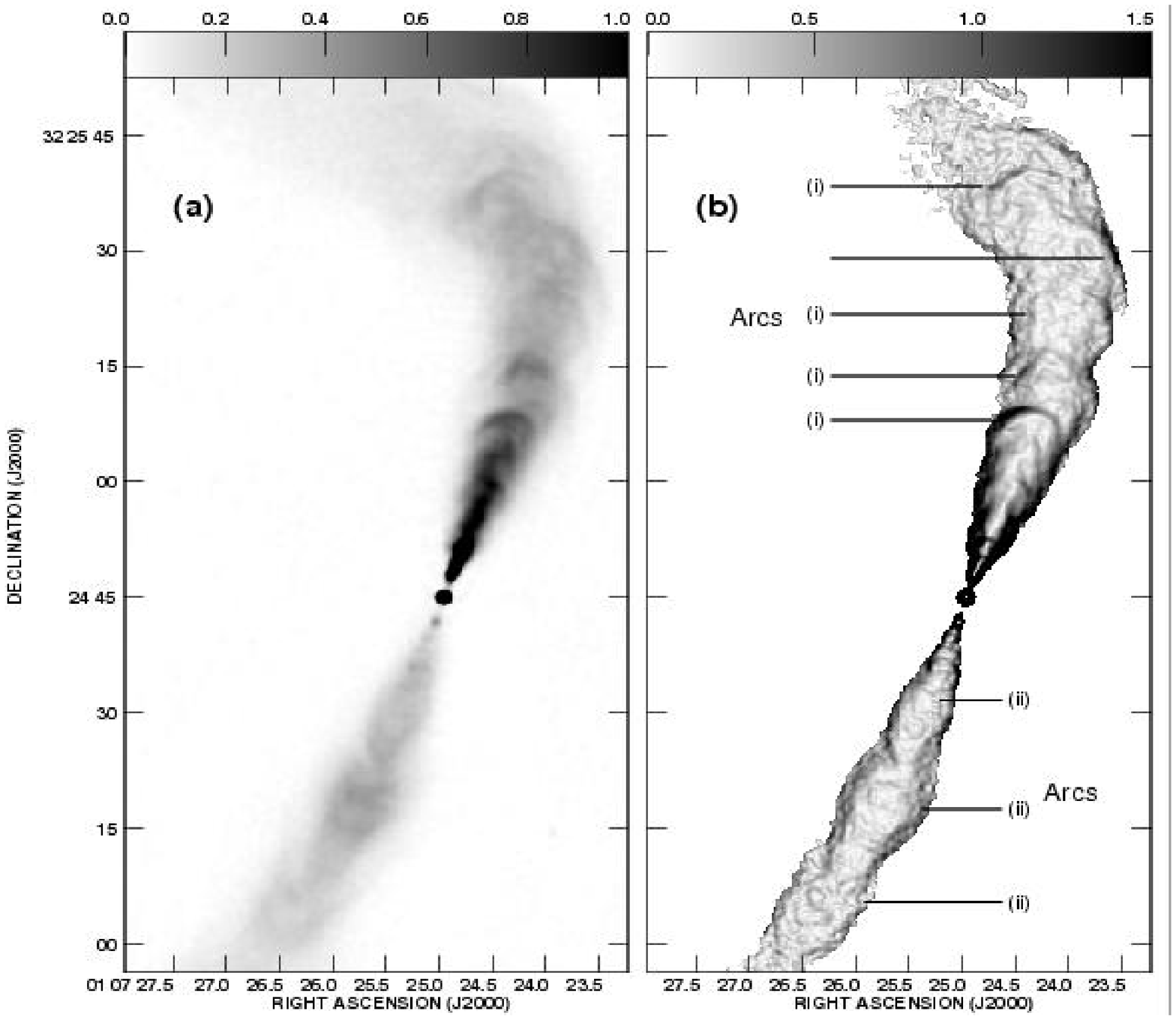}
\caption{Grey-scale image of 3C31 observed at 8440\,MHz with a resolution
of 0.75 arcsec FWHM.  (a) Total intensity (0 -- 1 mJy  beam$^{-1}$). (b)
Sobel-filtered and normalized by total intensity. The `arcs' discussed in
Section~\ref{Hi-res} are indicated and labelled with their morphological types
where these can be ascertained.\label{I0.75}}
\end{figure*}

The parameters of the images used in this paper are summarized in
Table~\ref{Image-table}.  All of the images have been corrected for the effects
of primary beam attenuation: as a consequence, the quoted rms noise levels
strictly refer to off-source regions close to the pointing centre.  We have used
the mean of the images at 1365, 1435, 1485 and 1636\,MHz to provide several of
the illustrations, and we refer to these as `mean L-band images' (using the
radio vernacular for designation of this frequency band).  We have checked the
registration of the images at different frequencies using the peak of the core
emission and isolated point sources in the field. We believe that the images at
1.5 and 5.5\,arcsec FWHM are aligned to better than FWHM/20.

\begin{figure*}
\epsfxsize=14cm
\epsffile{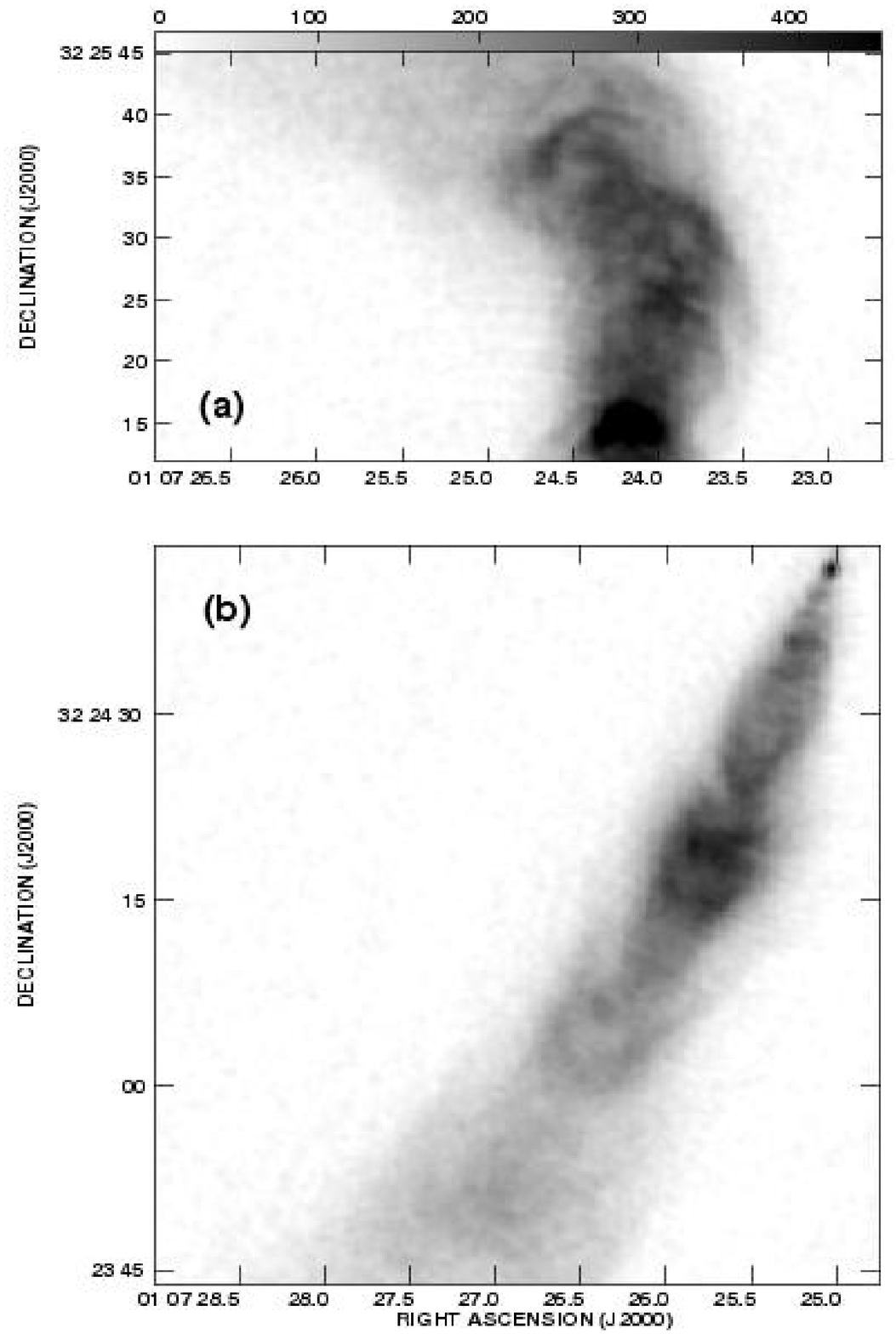}
\caption{Grey-scale images of 3C31 observed at 8440\,MHz with a resolution
of 0.75 arcsec FWHM.  The grey-scale range is 0 -- 0.5\,mJy beam$^{-1}$. 
(a) North jet, around the first bend; (b) South jet (note that the nucleus is
not included in this panel).\label{I0.75blowup}}
\end{figure*}

\begin{figure*}
\epsfxsize=17cm
\epsffile{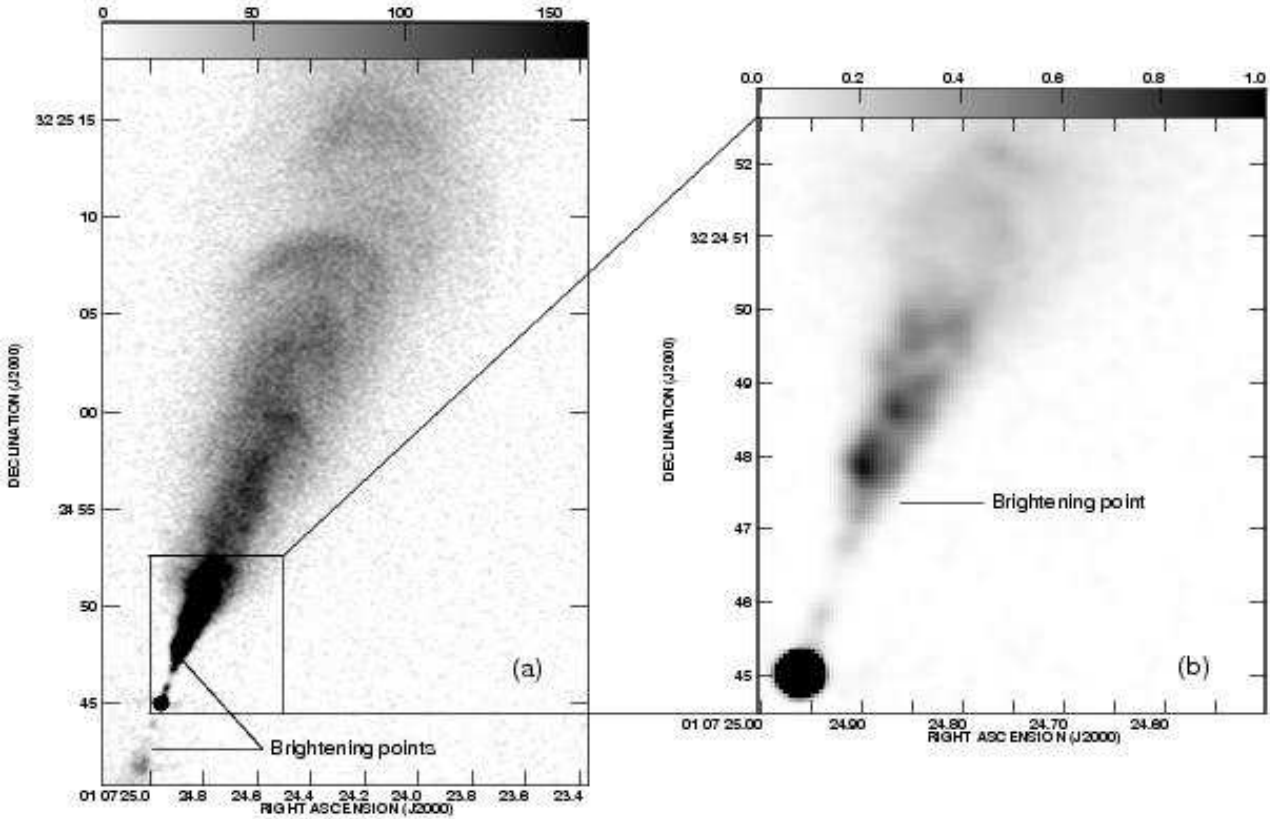}
\caption{Grey-scale images of 3C31 observed at 8440\,MHz with a resolution of
0.25 arcsec FWHM.  The grey-scale ranges for the two panels are: (a) 0 --
160\,$\mu$Jy beam$^{-1}$; (b) 0 -- 1\,mJy beam$^{-1}$. The box on panel
(a) shows the area covered by (b).\label{I0.25}}
\end{figure*}

\begin{figure}
\epsfxsize=8.5cm
\epsffile{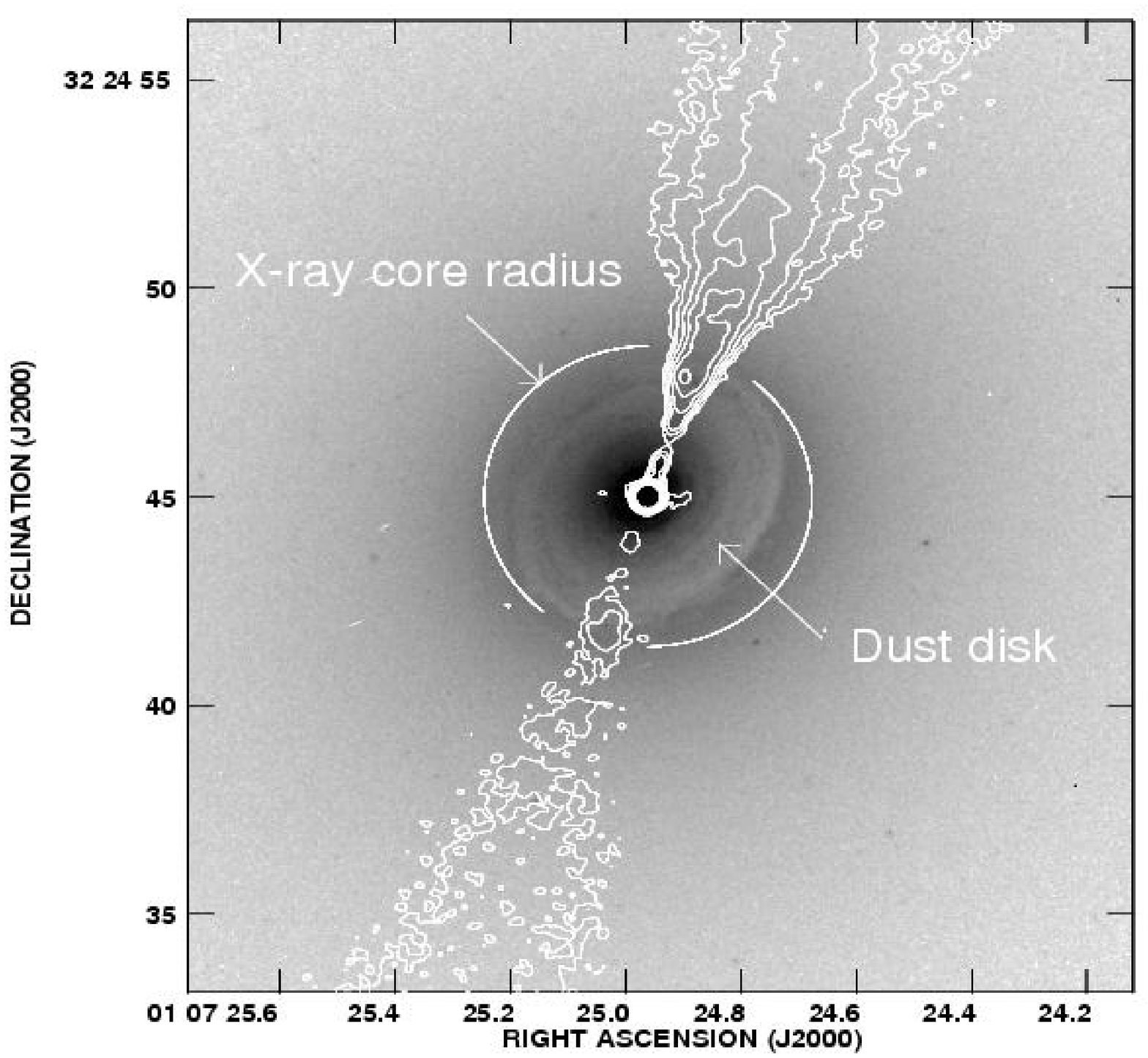}
\caption{Grey-scale of optical emission near 7000\AA\ observed with HST WFPC-2,
superimposed on contours of total intensity at 8440\,MHz with 0.25 arcsec 
FWHM resolution. The location of the dust disk and the core radius of the inner
hot gas component are marked \citep{Mar99,Hard02}. 
\label{Rad-HST}}
\end{figure}

The effects of limited short-spacing coverage and primary beam attenuation are
important at the two highest frequencies, particularly for spectral analysis. We
estimate that the 4985-MHz and 8440-MHz images are adequately sampled for scales
\la 300\,arcsec and \la 140\,arcsec, respectively, i.e.\ $\pm$150 and
$\pm$70\,arcsec from the nucleus \citep{ObsSS}.  There is less power on large
spatial scales in $Q$ and $U$, and we estimate that these are reliable at
significantly larger distances from the nucleus.  

Chromatic aberration (``bandwidth smearing'') causes point-like sources far
from the phase centre to appear distorted and reduced in peak flux
density. This effect is large at the periphery of our 5.5-arcsec L-band images 
(Fig.~\ref{Ideep}b) and typically causes a loss of peak intensity of 5 -- 10\%
for point sources at the edges of the areas we show in the remaining figures
\citep{ObsSS}.  Chromatic aberration preserves flux density, however, and
3C\,31 is fully resolved on all scales where it is significant. Its effect on
the images of the source shown in this paper is therefore very small compared
with other errors.  A 10-s integration time was used for all of the
observations, so time-average smearing is negligible over all of our fields
\citep{ObsSS}.

When deriving images of polarized power, $P = (Q^2+U^2)^{1/2}$, and degree
of polarization $p = P/I$, we made a first-order correction for Ricean
bias \citep{WK}.

\section{The images}
\label{Images}

\subsection{Large-scale structure}

Fig.~\ref{Ideep} shows the large-scale structure of 3C\,31.  The 40-arcsec
FWHM image (Fig.~\ref{Ideep}a) shows the maximum detectable extent of
the structure at high signal/noise ratio.  The lowest contour is
essentially identical to that seen in previous low-frequency images 
(\citealt{Burch77,Strom83,Jaegers,And92}). 

In Fig.~\ref{Ideep}(b), we show a grey-scale representation of the mean
L-band image at a resolution of 5.5\,arcsec FWHM, over a flux-density range chosen
to emphasise the low-brightness structure. All of the emission visible at
lower resolution is seen on this image, albeit at a lower level of
significance and with some artefacts.  Key regions of the source are
labelled: we refer to North and South {\em jets}, {\em spurs} and {\em tails}.
The essential distinction between tails and spurs is that the former
extend away from the nucleus whereas the latter extend back towards it, at
least in projection.  Filamentation and fine structure are already apparent
in Fig.~\ref{Ideep}(b): the bifurcation of the North tail and the filaments in
the North spur are best seen on this display.  The remarkably straight East edge
of the South spur is more prominent than on earlier images.

In a display of the mean L-band image at 5.5\,arcsec FWHM resolution over a
larger intensity range (Fig.~\ref{ISobel5.5}a), the true extent of the jets
becomes apparent. In order to display the intensity gradients more clearly, we
show a Sobel-filtered representation of the image in
Fig.~\ref{ISobel5.5}(b). The Sobel operator \citep{Pratt} computes $|\nabla I|$
and therefore enhances brightness gradients.  We show an image of fractional
intensity gradient $|\nabla I|/I$.  This image shows that the boundaries of the
jets are marked by strong fractional intensity gradients. These can be followed
to the positions marked N and S in Fig.~\ref{ISobel5.5}(b) and we therefore use
them to define the boundaries of the jet, spur and tail regions. At the points N
and S, there is evidence for curved features with sharp brightness gradients
slightly inside the outer boundary of the source. It is tempting to associate
these features with the redirection of the jets to form the tails and spurs. It
may also be that the spur and tail are parts of the same physical structure,
seen in projection on either side of the jet.

The South jet has a well-collimated region of comparatively low emissivity
(labelled `A' in Fig.~\ref{ISobel5.5}a) which begins where it becomes
superposed on the spur. Thereafter it bends through $\approx 90^\circ$ and
brightens before widening and appearing to merge with the spur and tail
emission. The North jet makes a sharp double bend before widening. Both jets
show complex, but well-resolved sub-structure after they widen. This
sub-structure is clearly evident in Fig.~\ref{ISobel5.5}(a), but is not
prominent in the fractional intensity gradient image
(Fig.~\ref{ISobel5.5}b).  The straight edge in the South spur noted earlier is
particularly prominent on the Sobel-filtered image
(Fig.~\ref{ISobel5.5}b).

\subsection{Total intensity images at 1.5 arcsec resolution}

A clearer representation of the termination of the jets is given by the
mean L-band image at 1.5-arcsec FWHM resolution (Fig.~\ref{I1.5}a), in which
the tail emission is heavily resolved. The internal structure of the jets
starts to be revealed by the 4985-MHz image at the same resolution
(Fig.~\ref{I1.5}b).  The main new features are as follows:
\begin{enumerate}
\item There is evidence for considerable filamentation and
other sub-structure within both jets. We discuss this in detail below
(Sections~\ref{Hi-res} and \ref{arc-pol}).  
\item The low-emissivity region A of the South jet mentioned earlier is seen in
more detail. It appears to narrow slightly before bending and brightening.
\item The region where the South jet expands inside the South tail shows complex,
resolved filamentary fine structure.
\end{enumerate}

\subsection{8440\,MHz total intensity images at 0.75 and 0.25 arcsec
resolution} 
\label{Hi-res}

\begin{figure}
\epsfxsize=8.5cm
\epsffile{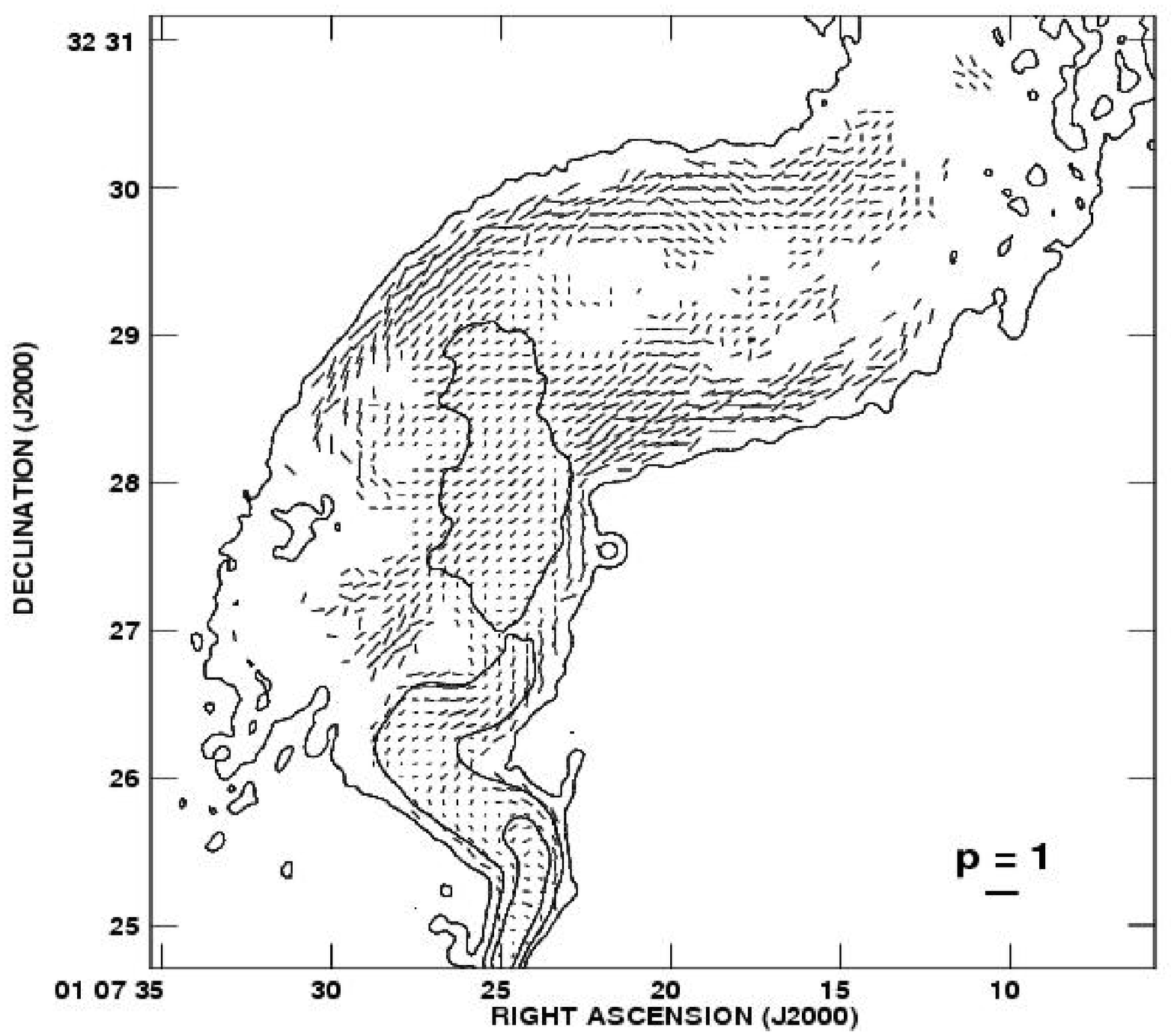}
\caption{Vectors whose lengths are proportional to the mean degree of
polarization, $p$, at L-band and whose position angles represent the apparent
magnetic field direction, $\chi(0) + \pi/2$, where $\chi(0)$ is the
zero-wavelength {\bf E}-vector position angle from a rotation-measure fit to all
four L-band frequencies \citep{lb08b}. The resolution is 5.5\,arcsec
FWHM and the polarization scale is indicated by the labelled bar.  The diagram
is designed to show the magnetic-field structure in the North tail over a region
which is not sampled by the 4985- and 8440-MHz
observations.\label{Bfield-low-N}}
\end{figure}

\begin{figure}
\epsfxsize=8cm
\epsffile{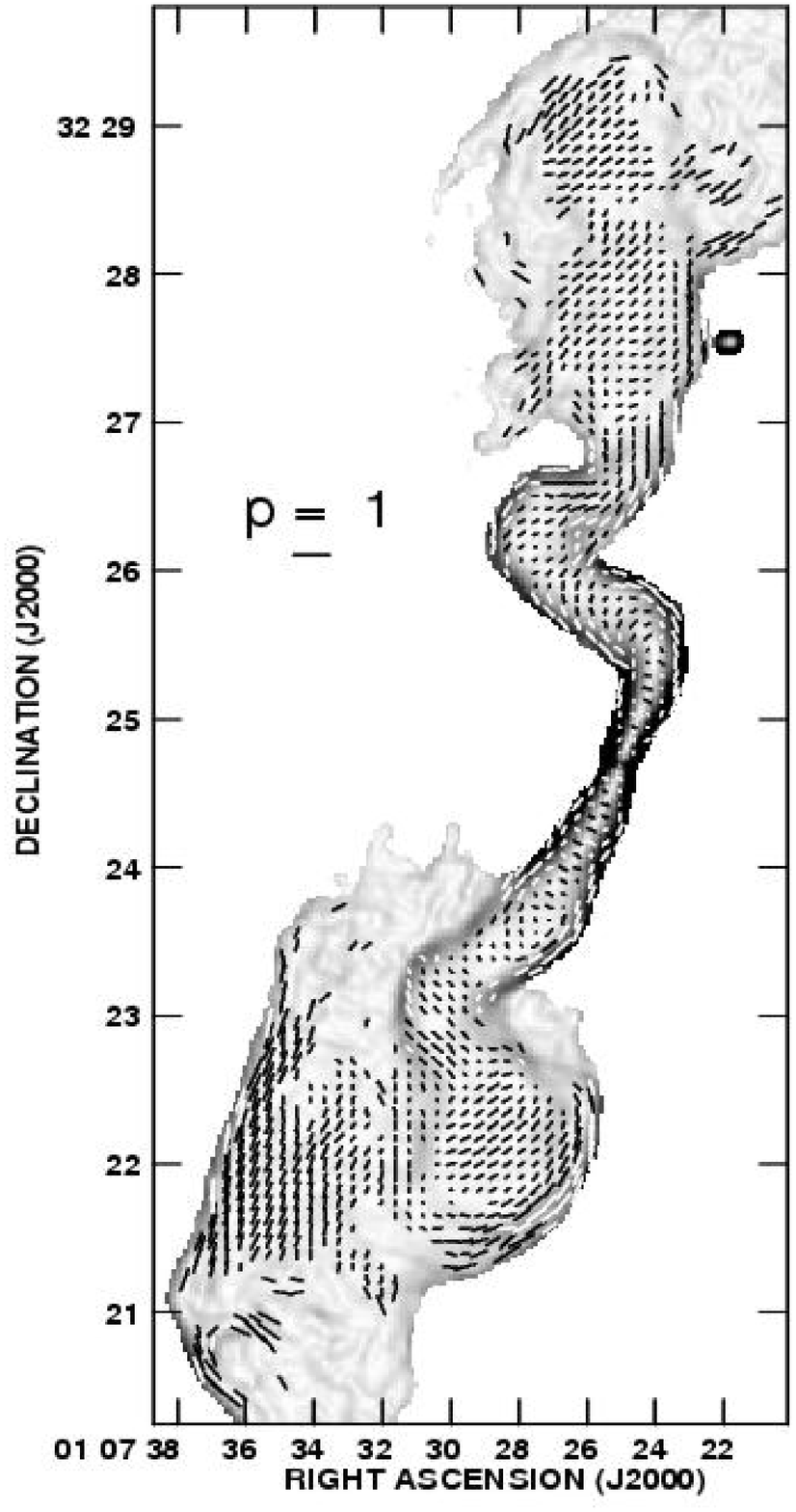}
\caption{Vectors whose lengths are proportional to the mean degree of
polarization, $p$, at 4985\,MHz and whose position angles represent the apparent
magnetic field direction, $\chi(0) + \pi/2$, where $\chi(0)$ is the
zero-wavelength {\bf E}-vector position angle from a rotation-measure fit to all
frequencies between 1365 and 4985\,MHz \citep[see][Fig.2c]{lb08b}. The
grey-scale is a Sobel-filtered, mean L-band image, normalized by total
intensity. The resolution is 5.5\,arcsec FWHM and the polarization scale is
indicated by the labelled bar.
\label{Bfield-low}}
\end{figure}

The higher-resolution images shown here cover a wider area than those discussed
by \citet{LB02a}.  The 0.75-arcsec FWHM image (Fig.~\ref{I0.75}a) shows a number
of features with sharp intensity gradients. Examples of these `arcs' are
labelled in Fig.~\ref{I0.75}(b) with their morphological types as defined by
\citet{LCBH06}. In order to display them more clearly,
Fig.~\ref{I0.75blowup} shows the relevant areas of the North and South jets on
an expanded scale and with a narrower grey-scale range.  The main
characteristics of the arcs are as follows.
\begin{enumerate}
\item The most prominent examples are at distances \ga 20\,arcsec (7\,kpc)
from the nucleus, but there are hints of similar structures closer in.
\item They all start from an edge of the jet and extend inwards towards
the jet axis in the downstream direction (i.e.\ away from the nucleus).
\item The most prominent examples are narrow and ridge-like, but
there are many fainter and broader features.
\item Three of the arcs in the North jet clearly cross the axis. They are
concave towards the nucleus and approximately semi-circular in form [the
defining characteristics of type (i) arcs given by \citet{LCBH06}].
\item The remaining arcs in the North jet have higher brightness contrast
at the outside edge of the bend. It is possible that they also extend
across most of the jet, but at lower contrast.
\item The features in the South jet have a greater degree of mirror symmetry
across the jet and show similar brightness contrast on both sides. They are type
(ii) arcs as defined by \citet{LCBH06}: oblique to the jet direction on either
side of the jet but without large brightness gradients on-axis. They are located
within the outer envelope of the jet emission, so the highest brightness
gradients do not occur at the jet edges (Fig.~\ref{I0.75blowup}b).
\item The maximum local brightness enhancement is $\approx$40\% (for the
brightest arc in the North jet); \la 20\% for the other arcs.  The implied
emissivity enhancements in the arcs would be of order 400\% to 200\%, {\it if} 
their depths along the line of sight are of the same order as their apparent 
widths.
\end{enumerate}
We discuss the magnetic-field structure and other properties of the arcs in
Section~\ref{arc-pol}. 

Finally, Fig.~\ref{I0.25} shows the inner  jet region at 0.25-arcsec
resolution in two different grey-scale representations. Fig.~\ref{I0.25}(a)
shows that the brightest arc in the North jet is fully resolved, with a
characteristic width $\approx$1\,arcsec.  The Figure also displays the knot at
the base of the counter-jet which marks the {\em brightening point} on that side
of the nucleus (we model the brightening as a sudden increase in emissivity at
2.5\,arcsec from the nucleus in projection on both sides; \citealt{LB02a}).
Fig.~\ref{I0.25}(b) illustrates the non-axisymmetric knot structure at the base
of the North jet.

\subsection{Optical-radio superposition at high resolution}

HST WFPC-2 observations of 3C\,31 were presented by \citet{Mar99}, who showed
that there is a complex central dust lane of radius $\approx$3 \,arcsec.  A
superposition of our 8440-MHz image at 0.25 arcsec FWHM resolution on the HST
image is shown in Fig.~\ref{Rad-HST}.  In order to correct for uncertainties in
the relative astrometry, we assumed that the brightest points on the optical and
radio images coincide.  The orientation and size of the kpc-scale rotating
molecular disk studied by \citet{Oku05} are consistent with those of the dust
lane. The radio jet flares and brightens on approximately the same (projected)
scale as significant substructure (banding) in the dusty molecular disk
(Fig.~\ref{Rad-HST}), but the latter is grossly misaligned with the jet axis and
there is no evidence for any {\it direct} interaction between the two structures
\citep{deK00}. The flaring scale is also comparable with the core radius of the
inner hot gas component identified by \citet{Hard02} and indicated on
Fig.~\ref{Rad-HST}. We have argued elsewhere that the flaring and recollimation
of the jets in 3C\,31 is governed by the pressure of hot gas associated with the
parent galaxy, rather than by interaction with the cool component
\citep{LB02b}.

\section{Field structure and intensity gradients}
\label{Morph}

We have estimated the degree of polarization at zero frequency, $p(0)$ and the
direction of the apparent magnetic field ($\chi(0) + \pi/2$, where $\chi(0)$ is
the zero-wavelength {\bf E}-vector position angle) by correcting the observed
values of $p$ and the {\bf E}-vector position angles for depolarization and
Faraday rotation, respectively.  The
depolarization in the North of the source at 5.5-arcsec resolution and over the
entire area visible at higher resolution is very small and the change of {\bf
E}-vector position angle is accurately proportional to $\lambda^2$, so these
corrections are well-determined \citep{lb08b}. The rotation-measure distribution and its
implications for the structure of the magnetoionic medium around 3C\,31 are
analysed in \citet{lb08b}; here we describe only the apparent field structure
within the jets and its relation to features of their total intensity.

\subsection{Field structure on large scales}
\label{Magnet-large}

Fig.~\ref{Bfield-low-N} shows the apparent magnetic-field structure of the outer
North jet, spur and tail derived from an L-band RM fit at
5.5\,arcsec FWHM \citep[see][Fig.~2b]{lb08b}. The degree of polarization is high
($p \approx$ 0.5 -- 0.6) at the edges of the North tail and spur. There is a
clear polarization minimum in the centre of the tail, suggesting that the
three-dimensional field configuration might be close to model B of \citet{L81},
where the longitudinal and toroidal components are roughly equal but there is no
field in the radial direction, perpendicular to the axis of the tail.

Fig.~\ref{Bfield-low} shows the apparent magnetic-field structure at 5.5-arcsec
resolution from a 5-frequency fit.  The remarkable straight edge of the South
spur is highly polarized at 4985\,MHz, with an apparent field along the edge,
suggesting that this part of the radio source has experienced compression or
shear at a large-scale planar boundary whose counterpart in the surrounding gas
has yet to be identified.

\begin{figure*}
\epsfxsize=17cm \epsffile{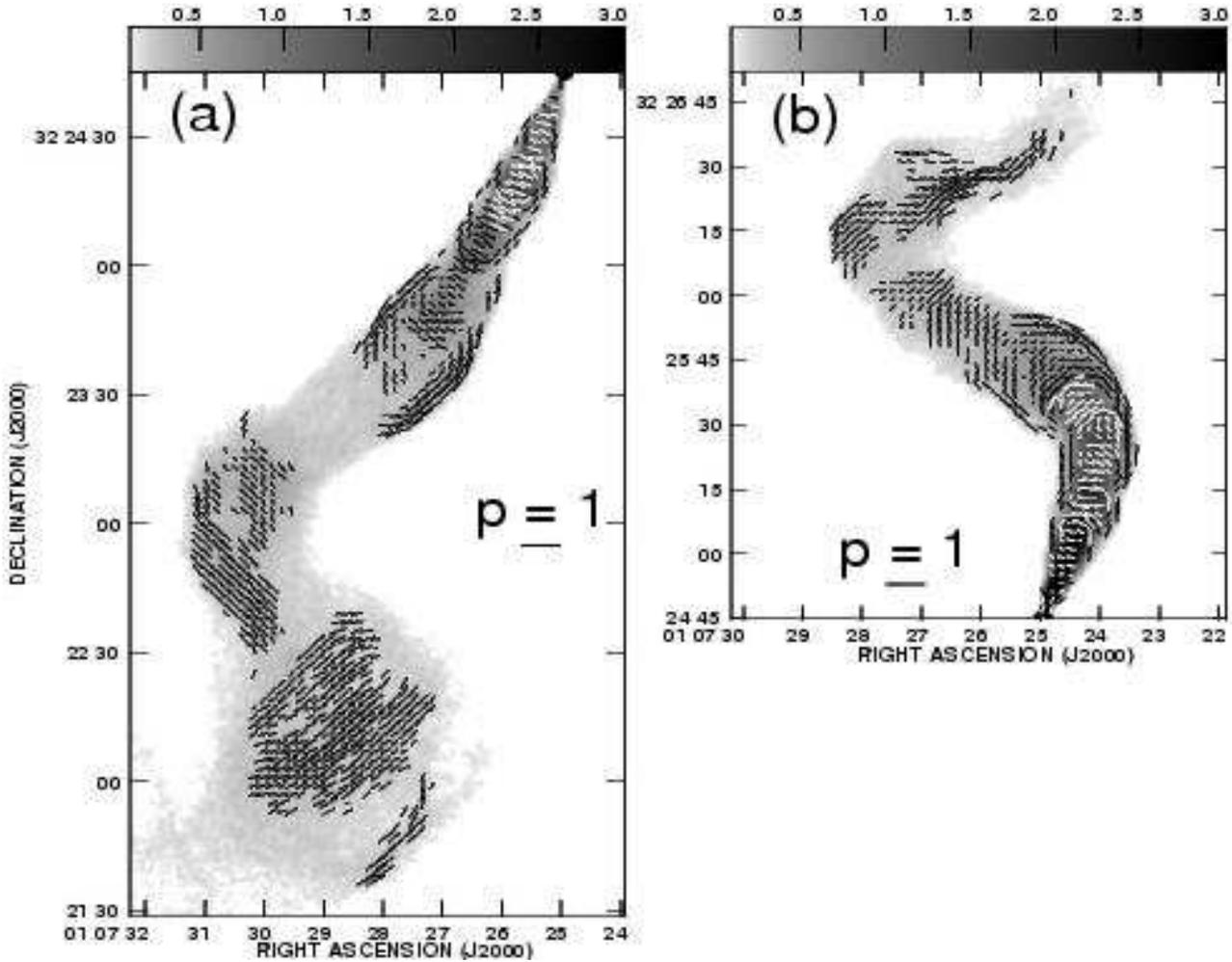}
\caption{Vectors whose lengths are proportional to $p(0)$, the degree of
polarization extrapolated to zero wavelength from a linear fit of $p$ against
$\lambda^2$ \citep{lb08b}. Their angles represent the apparent magnetic field
direction, $\chi(0) + \pi/2$, where $\chi(0)$ is the zero-wavelength {\bf
E}-vector position angle corresponding to the RM image of
\citet[Fig.~3c]{lb08b}. The vectors are superposed on a grey-scale of the
4985-MHz total intensity. The resolution is 1.5\,arcsec FWHM and the
polarization scale is indicated by the labelled bars. (a) South jet; (b) North
jet.
\label{Bfield1.5-fig}}
\end{figure*}

\begin{figure*}
\epsfxsize=17cm
\epsffile{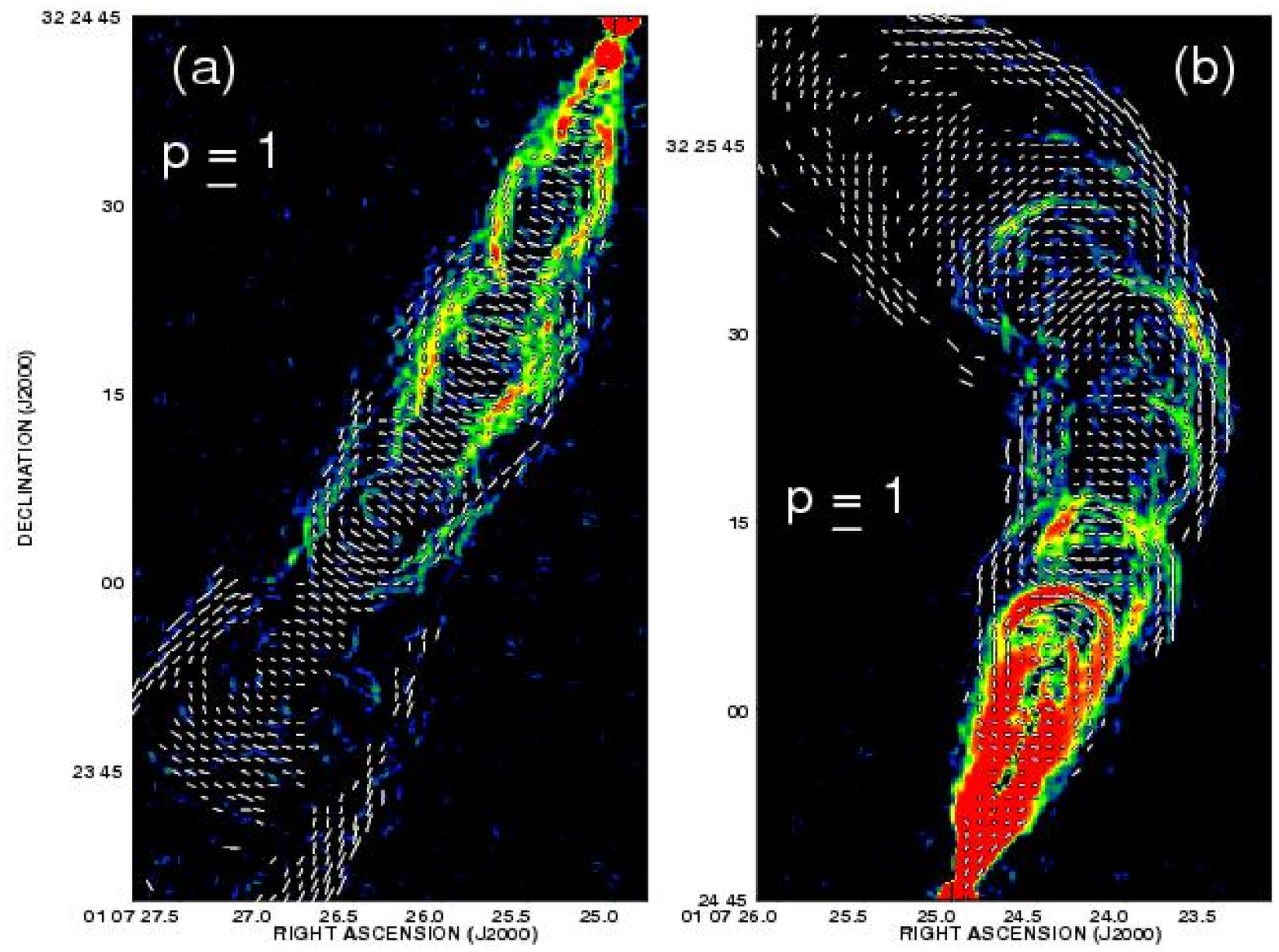}
\caption{The colour plots show Sobel-filtered, 8440-MHz $I$ images at a
resolution of 0.75\,arcsec FWHM superimposed on vectors whose magnitudes
are proportional to the degree of polarization, $p$, and whose directions
are those of the apparent magnetic field. The field directions were
derived by interpolating the RM image of \citet[Fig.~3c]{lb08b} onto a finer
grid and using it to correct the observed 8440-MHz {\bf E}-vector position
angles to zero wavelength. (a) South jet; (b) North jet.
\label{Bsobl-fig}}
\end{figure*}

At 1.5-arcsec resolution (Fig.~\ref{Bfield1.5-fig}), the apparent field
structure of the jets becomes clear.  To a first approximation, the outer
regions of both jets continue to show the configuration discussed in detail by
\citet{LB02a} for the inner 30\,arcsec -- an apparent field which is transverse
on-axis but longitudinal at the edges of the jets. In the North jet
(Fig.~\ref{Bfield1.5-fig}b) this configuration persists for at least another
30\,arcsec after the first bend. Thereafter, the on-axis field remains
transverse, but there is insufficient sensitivity to measure the polarization at
the jet edges.  After the second bend, the apparent field becomes longitudinal
over most of the jet.  The degree of polarization is significantly
enhanced at the outside edges of major bends in both jets.  Where the South jet
expands rapidly (Fig.~\ref{Bfield1.5-fig}a), it shows a predominantly transverse
apparent field, again with high $p$. There is a highly-polarized edge
where the jet makes its final sharp turn prior to disrupting at the feature S in
Fig.~\ref{ISobel5.5}. 

The apparent magnetic field in the brighter jet at 0.25-arcsec
resolution was shown by \citet{LB02a}.

\subsection{Magnetic fields in the arcs}
\label{arc-pol}

A superposition of the Sobel-filtered, 0.75-arcsec FWHM $I$ image and vectors
showing the magnitude of $p$ and the apparent field direction is shown in
Fig.~\ref{Bsobl-fig}. This reveals two striking effects:
\begin{enumerate}
\item The polarization is enhanced wherever there is a strong gradient in
total intensity, particularly at the arcs. For example,
$p \approx 0.4$ in the brightest arc in the North jet, even without correction
for contamination by surrounding emission.
\item The apparent field always appears parallel to the arcs. 
\end{enumerate}
The arcs in 3C\,296 show similar, but less prominent magnetic-field structures
\citep{LCBH06}.

In \citet{LCBH06}, we argued that the difference between type (i) arcs (found
predominantly in the main jets) and type (ii) (usually in the counter-jets) was
plausibly due to relativistic aberration. We suggested that a first-order model
of an arc is a thin, axisymmetric shell of enhanced emissivity, concave towards
the nucleus and travelling outward with approximately the velocity of the local
mean flow.  In that case, when our line of sight is tangential to part of the
shell {\it in its rest frame}, we will see enhanced emission.  Close to the
centre-line of the jet, the shell can be approximated as a planar sheet of
material orthogonal to the axis in the rest frame of the flow.  Relativistic
aberration causes the sheets to appear differently orientated to the line of sight
in the main and counter-jets so that those in the main jet can be observed
nearly edge-on while those in the counter-jet are always significantly rotated
about a line perpendicular to the jet axis in the plane of the sky and are
therefore less prominent both in intensity and in brightness gradient where they
cross the jet axis.  These differences between the main jet and counter-jet are
less pronounced for the slower flow at the edges of the jets, where the arcs can
therefore appear more symmetrical -- as observed.  If the magnetic field in the
arcs is compressed to lie in the plane of the shells, we also expect to see a
high degree of polarization with the apparent field direction along the arc, as
observed \citep{L80}.

The arcs therefore provide further
support for the hypothesis that the main differences in appearance between the
inner main jet and counter-jet in 3C\,31 arise from relativistic bulk motion in intrinsically 
similar jet outflows on the two sides of the nucleus. 
\section{Spectra}
\label{Spectrum}

\subsection{Comparison of images at different frequencies}

At 4985 and 8440\,MHz, we cannot observe the full range of spatial frequencies
covered by the D configuration at L band.  We have therefore investigated the
sensitivity of the spectral index determination between 1365 and 4985\,MHz at
5.5\,arcsec resolution to differences in the $(u,v)$ coverage by comparing the
spectral index distributions obtained from images made with:
\begin{enumerate} 
\item all of the observed data
at both frequencies, imaged with zero-spacing flux densities specified;
\item matched coverage at the two frequencies, with the central
portion of the $(u,v)$ plane, including the zero spacing, deleted at 1365\,MHz); 
\item as in (ii), but with the zero-spacing flux densities included.  
\end{enumerate} 
We show results only for areas of 3C\,31 where the spectral indices obtained by
these three methods agree ($\Delta\alpha < 0.1)$.  The limitations in $(u,v)$
coverage and primary beam size imply that we can derive reliable spectra only
for the North and South jets and the South spur between 1365 and 4985\,MHz at a
resolution of 5.5\,arcsec: the tails and the North spur are seen only at
L-band, within which the range of observed frequencies is too small for us to
derive reliable spectral indices.

\begin{figure*}
\epsfxsize=17cm
\epsffile{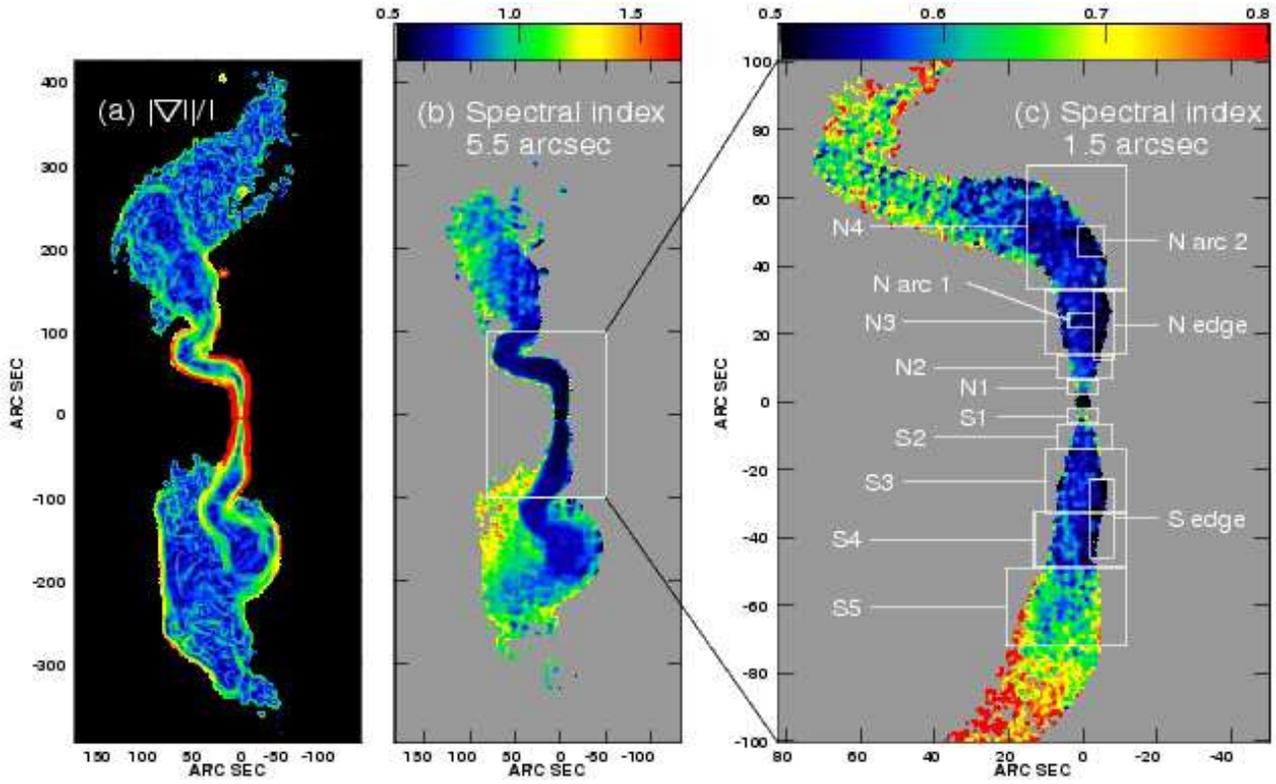}
\caption{(a) Sobel-filtered, mean L-band image (Section~\ref{Red}), normalized by total intensity,
  at a resolution of 5.5\,arcsec.  (b) and (c) Spectral index, $\alpha$ from
  weighted least-squares, power-law fits to the total intensity.  (b)
  5-frequency fit between 1365 and 4985\,MHz at a resolution of 5.5\,arcsec
  FWHM. Only points with $I > 5\sigma$ at all frequencies are shown.  The colour
  range is $0.5 \leq \alpha \leq 1.66$.  (c) 6-frequency fit to the
  maximum-entropy images at a resolution of 1.5\,arcsec FWHM for the inset area
  in panel (b). Only points with $I > 10\sigma$ at all frequencies were used in
  the fit.  The colour range is $0.5 \leq \alpha \leq 0.8$. The boxes mark the
  regions used for the fits in Fig.~\ref{specfits} and
  Table~\ref{spec-table}. All of the images have been rotated anticlockwise by
  19.7 deg.\label{Alpha-fig}}
\end{figure*}

Similar considerations apply at 1.5-arcsec resolution, where we have also
compared spectral-index distributions derived from  {\sc clean} and
maximum-entropy deconvolutions.  Spectra derived from the 8440-MHz images are
reliable only within $\pm$70\,arcsec of the nucleus, for the following reasons:
\begin{enumerate}
\item As mentioned earlier, the short-spacing coverage of the VLA
D-configuration is inadequate to image structure on larger scales
\citep{ObsSS}.
\item A comparison of spectral indices derived from images at 5.5 and
1.5-arcsec resolution, and from {\sc clean} and maximum-entropy images at
the higher resolution, showed differences at the level of $\Delta\alpha$
\ga 0.1 farther from the nucleus.
\item In order to keep the rms error in fitted spectral index $\la$0.1 at
1.5-arcsec resolution, we required that all input images had $I > 10\sigma$;
this condition was satisfied only in the central regions.
\end{enumerate} 
For the same reasons, we can search for deviations from a power-law spectrum 
only in the inner jets.

Experience suggests that the combination of calibration and deconvolution errors
results in an rms $\sigma_I \approx 0.03I$.  When fitting spectra, we have
therefore added this term and the off-source noise from Table~\ref{Image-table}
(corrected for primary-beam attenuation) in quadrature. As noted in
Section~\ref{Red}, chromatic aberration does not affect the measured
brightnesses for 3C\,31 significantly. The reason is that the effect preserves
flux density and 3C\,31 is very well resolved on all scales where it is
important.

\subsection{Power-law spectra}

The best representation of the spectrum at a resolution of 5.5\,arcsec is given
by a weighted least-squares fit of power-law spectra to the $I$ images over the
frequency range 1365 -- 4985\,MHz (Fig.~\ref{Alpha-fig}b). The `spectral
tomography' technique of \citet{KSR97} and \citet{KS99} leads to similar
conclusions, but our use of all of the available frequencies minimizes noise.

\begin{figure}
\epsfxsize=8.5cm
\epsffile{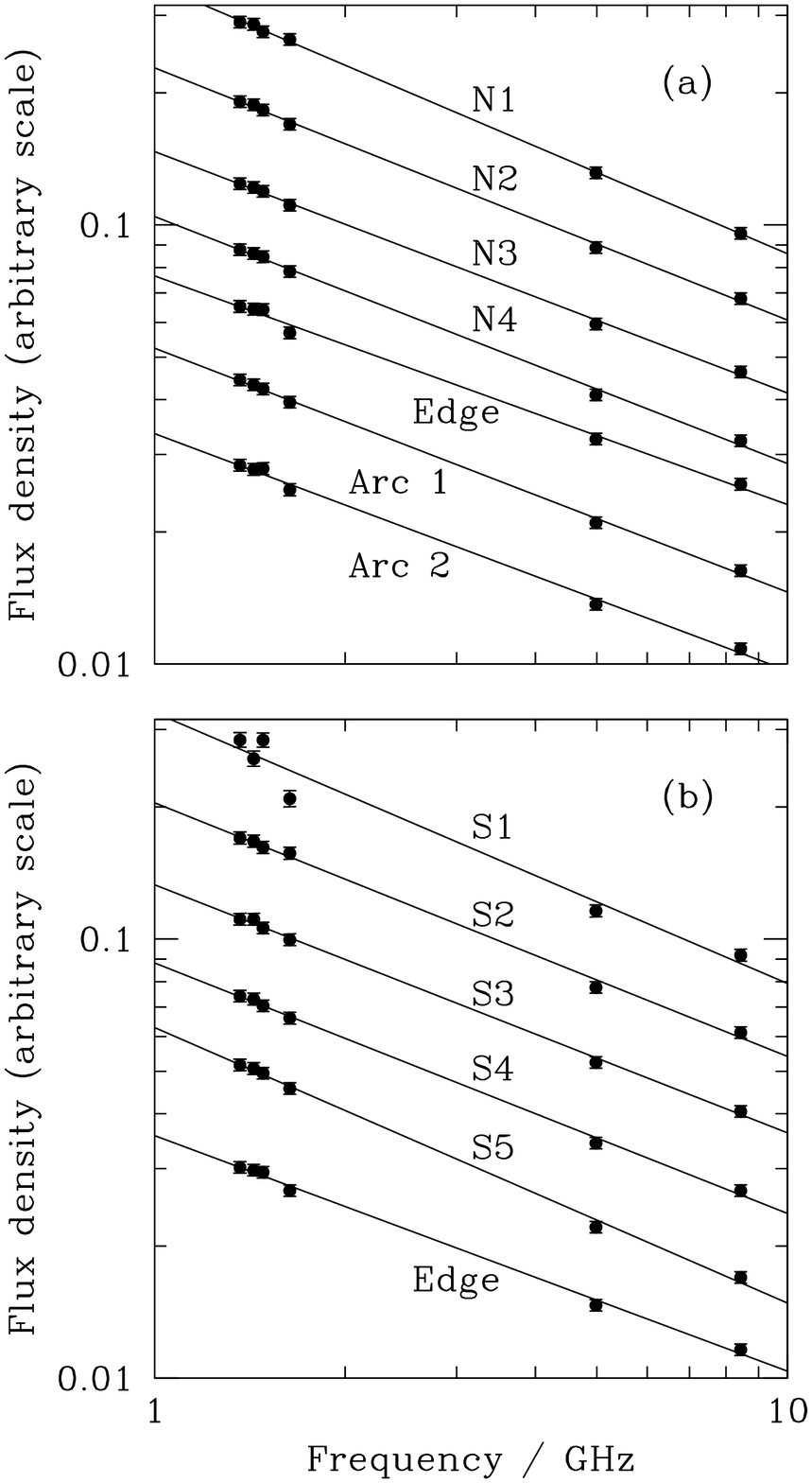}
\caption{Power-law fits to the total-intensity images over the rectangular
regions shown in Fig.~\ref{Alpha-fig}(c). (a) North jet, (b) South
jet.\label{specfits}}
\end{figure}

The spectral index variation over the inner jets is shown in more detail in
Fig.~\ref{Alpha-fig}(c), which is derived from a six-frequency power-law fit at
1.5-arcsec resolution.  Points are only plotted if: (a) $I > 10\sigma$ at all
frequencies (resulting in a maximum rms error in spectral index of 0.11) and (b)
the spectral indices derived from {\sc clean} and maximum-entropy images differ
by $<$0.075.  In order to search for deviations from power-law spectra at high
signal-to-noise ratio, we have integrated the flux densities over the
rectangular regions shown in Fig.~\ref{Alpha-fig}(c) for the 1.5-arcsec FWHM
images, using the same blanking criterion as for the spectral-index map in
Fig.~\ref{Alpha-fig}(c). The results are shown in Fig~\ref{specfits} and the
spectral indices are tabulated in Table~\ref{spec-table}.

\subsection{Spectral variations}

\subsubsection{Description}

The principal features of the spectra shown by Figs~\ref{Alpha-fig} and
\ref{specfits} are as follows:
\begin{enumerate}
\item There is no evidence for any significant deviation from power-law spectra
wherever we have reliable 8440-MHz data (i.e.\ within 70\,arcsec of the
nucleus).  All of the fits shown in Fig.~\ref{specfits} except for S1 have a
probability $>0.45$ that the chi-squared statistic would exceed the measured
value by chance, suggesting that we have slightly overestimated the
multiplicative error term. For S1 (the faint region of the counter-jet close to
the core) there are significant residual artefacts at L band which are not
represented in our noise estimate.  Our constraints on spectral curvature are
limited by calibration uncertainties at the 1 -- 2\% level (it is clear from
Fig.~\ref{specfits} that the 1636- and 4985-MHz points are slightly low relative
to the best-fitting power law and the 8440-MHz points slightly high). In any
case, the curvature expected over our restricted frequency range is small,
because of the finite width of the synchrotron emission spectrum for a single
electron. For these reasons, we will not discuss spectral curvature in what
follows.  As the calibration differences are common to all of the fits, the
differences in spectral index {\it between regions} are better determined than
the absolute values or curvature, however, with an estimated rms error of only
0.01.
\item Within $\approx$7\,arcsec of the nucleus in both jets, the spectral index
  at 1.5-arcsec resolution is slightly steeper ($\langle \alpha \rangle = 0.62$)
  than the average for the inner jets.
\item From 7 -- 50\,arcsec in both jets, the mean spectral index is in the range
  0.55 -- 0.57. Farther from the nucleus, there is a gradual spectral steepening. 
\item There are also slight, but significant variations in spectral index across
both jets  within $\approx$30\,arcsec of the nucleus in
the sense that their West edges tend to have flatter spectra ($\langle \alpha
\rangle = $ 0.52 -- 0.54; Fig.~\ref{Alpha-fig}c).  We are confident that this is not an
instrumental effect: it is clearly visible at both resolutions and on the {\sc
clean} and maximum-entropy images.  Any residual misalignments of the images,
particularly at 1.5-arcsec resolution, should be extremely small. The most
likely remaining error (confusion with the unresolved radio core at the base of
the North jet) would not produce a difference in spectral index across the jets.
Flatter spectra are also seen on both edges of the jets in NGC\,315
\citep{LCCB06}.
\item The spectra of the two brightest arcs in the North jet are also 
slightly flatter than that of the surrounding emission
(Fig.~\ref{Alpha-fig}c; Table~\ref{spec-table}).
\item There is a clear spectral separation between the South jet and the spur
emission, matching the separation of these regions defined by the sharpest
brightness gradients (Figs~\ref{Alpha-fig}a and b).  This is particularly clear
where the jet first enters the diffuse emission. The spectral identity of the
jet is evidently maintained even after it bends abruptly about 2\,arcmin South
of the nucleus and remains until it terminates in a region of high brightness
gradient.  The outer South jet in 3C\,31 is therefore a clear example of the
type of spectral structure noted in other FR\,I sources \citep{KSR97,KS99}
wherein a flatter-spectrum `jet' with a distinct spectral identity is superposed
on steeper-spectrum `sheath' emission.
\item There are hints of a similar spectral difference between the jet and
the spur and tail in the North, but the sensitivity and short-spacing
coverage of the 4985-MHz observations are inadequate to confirm this. The
difference is seen clearly in WSRT observations at 0.6 and 1.4\,GHz
with a resolution of 29 $\times$ 55\,arcsec FWHM (fig.~3g of
\citealt{Jaegers}).
\item Approximate subtraction of the steeper-spectrum sheath emission by
  linear interpolation under the jets shows that there must be some intrinsic
  steepening of the jet spectrum with increasing distance from the nucleus in
  addition to the apparent steepening caused by emission from sheath material
  along the line of sight.  The complex morphology of the South spur and tail
  make it difficult to separate the spectral components unambiguously, however.
\item At 5.5-arcsec resolution, the spectral index of the North jet is flatter
by 0.1 -- 0.2 on-axis than at the edges in the region of the double bend,
between 80 and 120\,arcsec from the nucleus, the flattest spectrum being
associated with a bright filament (Fig.~\ref{Alpha-fig}b).  Similarly, transverse
spectral gradients in the opposite sense to those described in (iv) -- i.e.\
with the jet spectrum flatter on-axis -- are seen at distances $\ga$50\,arcsec
from the nucleus at 1.5-arcsec resolution (Fig.~\ref{Alpha-fig}c). These effects
are most likely to be produced by blending of flat-spectrum emission from the
jets with steep-spectrum emission from the surrounding spurs and tails.
\item \citet{And92} found that within 9\,arcmin of the core the spectrum
observed with low resolution flattens 
above 5\,GHz, as previously suspected by \citet{Burch77}. They interpreted
this effect as a superposition of components with different spectra, a
conclusion which is confirmed directly by our data. 
\end{enumerate}

\begin{table}
\caption{Spectral indices from 6-frequency power-law fits for the regions shown
  in Fig.~\ref{Alpha-fig}(c). The flux densities used in the fits were derived
  by integration over the images at 1.5-arcsec resolution. The rms errors on the
  spectral indices are all between 0.017 and 0.020, and are dominated by
  calibration uncertainties.  The rms errors on spectral-index differences
  between regions are $\approx$0.01. \label{spec-table}}
\begin{tabular}{lc}
\hline
&\\
Region & $\alpha$ \\
&\\
N arc 2&0.54  \\
N arc 1&0.56  \\
N edge &0.52  \\
N4     &0.56  \\
N3     &0.55  \\
N2     &0.58  \\
N1     &0.62  \\
S1     &0.62  \\
S2     &0.58  \\
S3     &0.57  \\
S4     &0.57  \\
S5     &0.63  \\
S edge &0.54  \\
&\\
\hline
\end{tabular}
\end{table}

\subsubsection{Spectral variations in FR\,I jets}

These observations contribute to the developing picture of spectral variations
in the bases of FR\,I jets, as follows.
\begin{enumerate}
\item Where the jets first brighten, there is a remarkably small dispersion
 around a spectral index of $\alpha = 0.62$ in the three sources we have studied
 in detail: 3C\,31, NGC\,315 \citep{LCCB06} and 3C\,296 \citep{LCBH06}, as well
 as 3C\,66B \citep{HBW}.  The average is dominated by emission immediately after
 the brightening point (Fig.~\ref{I0.25}).
\item The spectral index of the fainter emission close to the nucleus in 3C\,449
 \citep{RKS97}, PKS1333$-$33 \citep{KBE} and 3C\,66B \citep{HBW} appears to be
 comparable or slightly steeper although the uncertainties are larger. We do not
 have adequate resolution to measure the spectral index in these parts of the jets
 in 3C\,31.
\item Farther from the nucleus, the radio spectra flatten slightly to $\alpha = $ 0.50
 -- 0.55. This flattening  occurs where
 our kinematic models require deceleration \citep{LB02a,CLBC,LCBH06}. 
\item A related result is that an asymptotic low-frequency spectral index of
  0.55 is common in FR\,I jets over larger areas than we consider here
  \citep{Young}.
\item Flatter-spectrum edges can be seen where the jets are isolated from
  significant surrounding diffuse emission. Our kinematic models
  \citep{LB02a,CLBC,LCBH06} show that all of the jets have substantial
  transverse velocity gradients in these regions.
\item Our kinematic models and spectral measurements together suggest a
  relation between spectral index and flow speed $\beta = v/c$, ranging from
  $\alpha = 0.62$ for $\beta \approx 0.8$ to $\alpha \approx 0.5$ for $\beta \la
  0.2$.  Spectra would then be expected to flatten with distance from the
  nucleus (as the jets decelerate) and from centre to edge (as a result of
  transverse velocity gradients).
\item Synchrotron X-ray emission from the main jets in 3C\,66B, 3C\,31, 3C\,296
 and NGC\,315 is strongest relative to the radio close to the nucleus, at or
 before the brightening point \citep{HBW,Hard02,Hard05,Wor07}.  X-ray emission
 is still detected from the flatter-spectrum regions farther out, but at a lower
 level relative to the radio.
\item There is approximate morphological correspondence between the radio and
  X-ray brightness distributions, but they differ in detail. In the
  best-resolved case, NGC\,315 \citep{Wor07}, the X-ray emission is clearly
  extended across the jet.  Particle acceleration appears to be distributed
  throughout the jet volume, rather than being exclusively associated with
  discrete knots or with the boundary.
\item The ratio of X-ray to radio emission decreases and the radio spectrum
  starts to flatten where our kinematic models show that the jets decelerate
  rapidly from speeds of $\beta \approx$ 0.8 -- 0.9 \citep{LB02a,Hard02,CLBC,Wor07}.
\item The strongest X-ray emission is not associated with the flattest radio
 spectra, but rather with some particle acceleration process whose
 characteristic energy index is $s = 2\alpha + 1 = 2.24$. This is intriguingly
 close to the asymptotic value of $s = 2.23$ for first-order Fermi acceleration
 at relativistic shocks in the limit of large shock (or, equivalently, upstream)
 Lorentz factor $\Gamma$, a result which holds even for relativistically hot
 jets (\citealt{Kirk2000} and references therein; \citealt{Kirk2005}). The
 asymptotic value of $s$ is only approached for $\Gamma = (1-\beta^2)^{-1/2}
 \ga$10 \citep{Kirk2000,LP03}, however, so we would have to suppose that some
 fraction of the flow before the brightening point has a bulk Lorentz factor
 $\Gamma \ga 10$ and decelerates to $\Gamma \approx 2$, as required by our
 kinematic models. The shock would then have to be oblique, since the normal
 component of the the incoming plasma speed would be reduced to $\beta \approx
 1/3$ at the shock. Such a flow would be difficult to detect because of the
 narrow beaming angle, even if it contains relativistic particles.  Emission
 before the brightening point would then come primarily from a slow surface
 layer, as suggested for 3C\,31 by \citet{LB02a}.
\item As the jet slows down and the shocks become both less relativistic and
  weaker, the energy index would decrease, eventually reaching $s = 2$ ($\alpha
  = 0.5$) when the shocks become non-relativistic \citep{Bell78}.  Although this
  provides a natural reason for the spectrum of a jet to flatten as it
  decelerates, the required flow speeds are higher than we infer. In particular, 
  the spectral index of $\alpha = 0.61$ is seen over a distance of
  $\approx$8\,kpc (corrected for projection) along the jet axis in NGC\,315, in
  a region where we infer the on-axis flow to have $\Gamma \approx 2.1$
  \citep{CLBC,LCCB06}.  We would then have to suppose that some part of the flow
  remains ultrarelativistic (and effectively invisible) on large scales. 
\item A second possibility is that the process
  that produces the flatter spectrum is associated with shear \citep{LCCB06}.
  In 3C\,31, flatter-spectrum edges occur predominantly on the Western edges of
  the jets, i.e.\ on the outer edges of bends, perhaps consistent with this
  idea.  Second-order Fermi acceleration driven by turbulence in the shear layer
  \citep{SO} appears to be a viable process in FR\,I jets, but the shear
  acceleration mechanism described by \citet{RD} is unlikely to be efficient
  enough to accelerate the X-ray-emitting electrons in the relatively modest
  velocity gradients we infer.
\item The idea that two different acceleration processes are required has also
  been suggested on the basis of evidence from the X-ray morphology and
  spectrum of the brighter jet in Cen\,A \citep{K06,Hard07}. Here, the compact
  knots observed close to the nucleus are thought to be associated with shocks,
  while a truly diffuse acceleration mechanism dominates at larger
  distances.  
\end{enumerate}

Far from the nucleus, the picture is complicated by the presence of diffuse
emission surrounding the jets. As well as a smooth steepening of the jet
spectrum (expected from synchrotron and adiabatic losses affecting a homogeneous
electron population), multiple spectral components are observed. Jets appear to
retain their identities even after entering regions of diffuse emission and are
clearly identifiable by their flatter spectra. They are usually separated from
the surrounding emission by sharp brightness gradients.  This 
separation is observed in FR\,I sources with bridges of emission extending back
towards the nucleus (e.g.\ 3C\,296; \citealt{LCBH06}) as well as sources with
tails and spurs like 3C\,31. Although there is an overall trend for the spectrum
of the diffuse emission to steepen towards the nucleus in bridges and away from
it in tails \citep{Parma99}, the variations in individual objects such as 3C\,31
are complex.  The termination regions of jets in tailed FR\,I sources are
perhaps best regarded as bubbles which are continually fed with fresh
relativistic plasma by the jets and which in turn leak material into the tails.
Their spectral steepening would then be governed by a combination of continuous
injection, adiabatic, synchrotron and inverse Compton energy losses and escape.

\section{Summary and Conclusions}
\label{Conclusions}

Our new images of the FR\,I radio galaxy 3C\,31 reveal a complex substructure of
arcs and filaments within the jets and tails that is clearly related to the
organization of the apparent magnetic field structure. There is a strong
correspondence between steep brightness gradients and the organization of the
apparent field direction perpendicular to such gradients.

\citet{LB02a} have shown that the detailed brightness and polarization
properties of the inner jets in 3C\,31 are consistent with interpreting them as
intrinsically symmetrical relativistic flows that decelerate with increasing
distance from the nucleus, so that the observed asymmetries also decline with
distance.  We suggest that the observed differences between the intensity and
polarization structures of the arcs in the inner main and counter-jets are also
caused by the effects of aberration on emission from intrinsically similar
structures.

Our new data further suggest that we are observing the transition between
predominantly relativistic asymmetries which dominate in the straight inner
parts of the main jet and the counter-jet and environmental asymmetries which
dictate the appearance of 3C\,31 at distances $\ga$1\,arcmin from the nucleus.
Effects of an asymmetric environment include: (a) the different scales on which
bending begins in the North and South jets; (b) the dissimilar morphologies of
the North and South spurs and tails, and (c) the lack of any systematic
correspondence between the locations, shapes and brightnesses of substructures 
in the outer parts of the North and South jets.

Our images also show that both jets in 3C\,31 retain their spectral identity far
from the galactic nucleus, where their emission is superimposed on larger-scale,
steeper- spectrum structures.  Spectral imaging and intensity-gradient imaging
(Sobel filtering) techniques both delineate the {\it same} boundaries between the
jets and their environs on both the North and South sides of the galaxy.

The radio spectra of the inner jets of 3C\,31 and other FR\,I radio galaxies
exhibit systematic trends which are related both to jet kinematics and to X-ray
emission:
\begin{enumerate}
\item The characteristic radio spectral index where the jets first brighten is
  $\alpha = 0.62 \pm 0.01$. The ratio of X-ray/radio emission is highest at and
  before the brightening point.
\item Farther out, the spectrum flattens slightly, to $\alpha \approx 0.50 -
  0.55$ and the X-ray emission is fainter relative to the radio. The onset of
  this spectral flattening coincides with deceleration of the flow in our
  kinematic models.
\item Flatter spectra are seen at the jet edges where the jets are isolated
  from significant diffuse emission, in regions where our kinematic models
  require significant transverse velocity gradients.
\end{enumerate}
The correlation between radio spectral index and X-ray emission suggests a
direct association with the particle-acceleration process. The energy index of
2.24 corresponding to $\alpha = 0.62$ is close to the asymptotic value for
relativistic shocks in the limit of high Lorentz factors, and the spectral
flattening could result from progressive weakening of the shocks as the jets
slow down. The Lorentz factors of at least some parts of the flow would have to
be much higher than we infer, however. Alternatively, there may be two different
acceleration mechanisms operating on kpc scales in these jets, one dominant from
the nucleus until the region where jets brighten abruptly in the radio, the
other taking over at larger distances and perhaps associated with transverse
velocity shear.
 
\section*{Acknowledgements}

RAL would like to thank the Istituto di Radioastronomia, NRAO and Alan and Mary
Bridle for hospitality during the course of this work.  We acknowledge travel
support from NATO Grant CRG931498. We thank John Kirk for advice on particle
acceleration mechanisms and the referee, John Wardle, for his helpful report.
The National Radio Astronomy Observatory is a facility of the National Science
Foundation operated under cooperative agreement by Associated Universities, Inc.

\end{document}